\documentclass[sn-mathphys]{sn-jnl}

\usepackage{amssymb}
\usepackage{graphicx}
\usepackage{epstopdf}
\usepackage{amsmath}
\usepackage{epsfig}
\usepackage{color}
\usepackage{ulem}
\usepackage{stmaryrd}

\jyear{2021}

\theoremstyle{thmstyleone}

\theoremstyle{thmstyletwo}

\theoremstyle{thmstylethree}

\begin{document}

\title[ Deciphering Complexity: Machine Learning Insights into Chaotic Dynamical Systems ]{Deciphering Complexity: Machine Learning Insights into Chaotic Dynamical Systems}

\author*[1]{\fnm{Lazare} \sur{Osmanov}}\email{lazare.osmanov1521r@gmail.com}

\affil*[1]{\orgdiv{School of Physics}, \orgname{Free University of Tbilisi}, \orgaddress{\street{David Aghmashenebeli Alley}, \city{Tbilisi}, \postcode{0159}, \country{Georgia}}}

\abstract{

We introduce new machine-learning techniques for analyzing chaotic dynamical systems. The primary objectives of the study include the development of a new and simple method for calculating the Lyapunov exponent using only two trajectory data points unlike traditional methods that require an averaging procedure, the exploration of phase transition graphs from regular periodic to chaotic dynamics to identify "almost integrable" trajectories where conserved quantities deviate from whole numbers, and the identification of "integrable regions" within chaotic trajectories. These methods are applied and tested on two dynamical systems: "Two objects moving on a rod" and the "Henon-Heiles" systems.

}


\maketitle

\section{Introduction}\label{sec1}

In recent years, artificial intelligence (AI), machine learning, and deep learning have made significant advancements across various scientific disciplines. These techniques have permeated fields such as physics, where they find application in laboratory experiments in high energy and condensed matter physics and astronomical observations \cite{a1,a2,a3,a4,a5}.
This paper focuses on applying machine learning methods to chaos theory within dynamical systems.

\begin{figure}
\resizebox{\hsize}{!}{\includegraphics[width=10cm]{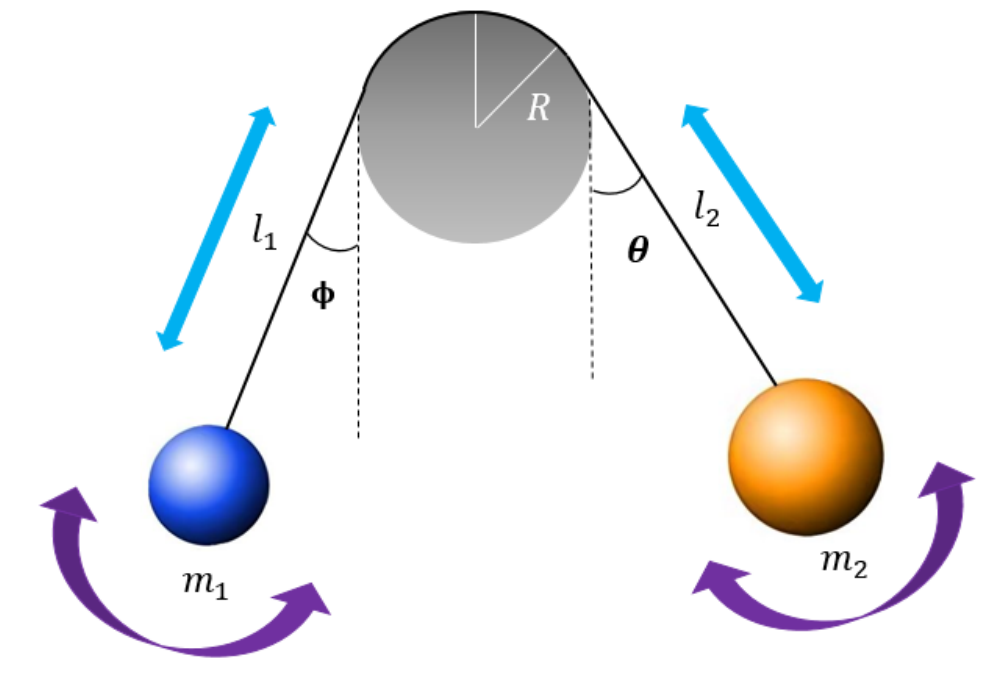}}
\caption{Schematic picture from \cite{me} for the "two bodies swinging on a rod" system. $m_1$, $m_2$ are masses of the loads that are attached to the ropes ends, $l_1$, $l_2$ are distances between the loads and touching points of the rope with the rod. The radius of the rod is $R$ and the full length of the rope is $L$. The rod is a frictionless surface. The arrows indicate oscillation directions of the loads,  angle $\theta$ grows in the positive, counterclockwise direction, while $\phi$ in the clockwise direction.} \label{fig5}
\end{figure}

\begin{figure}
\resizebox{\hsize}{!}{\includegraphics[width=10cm]{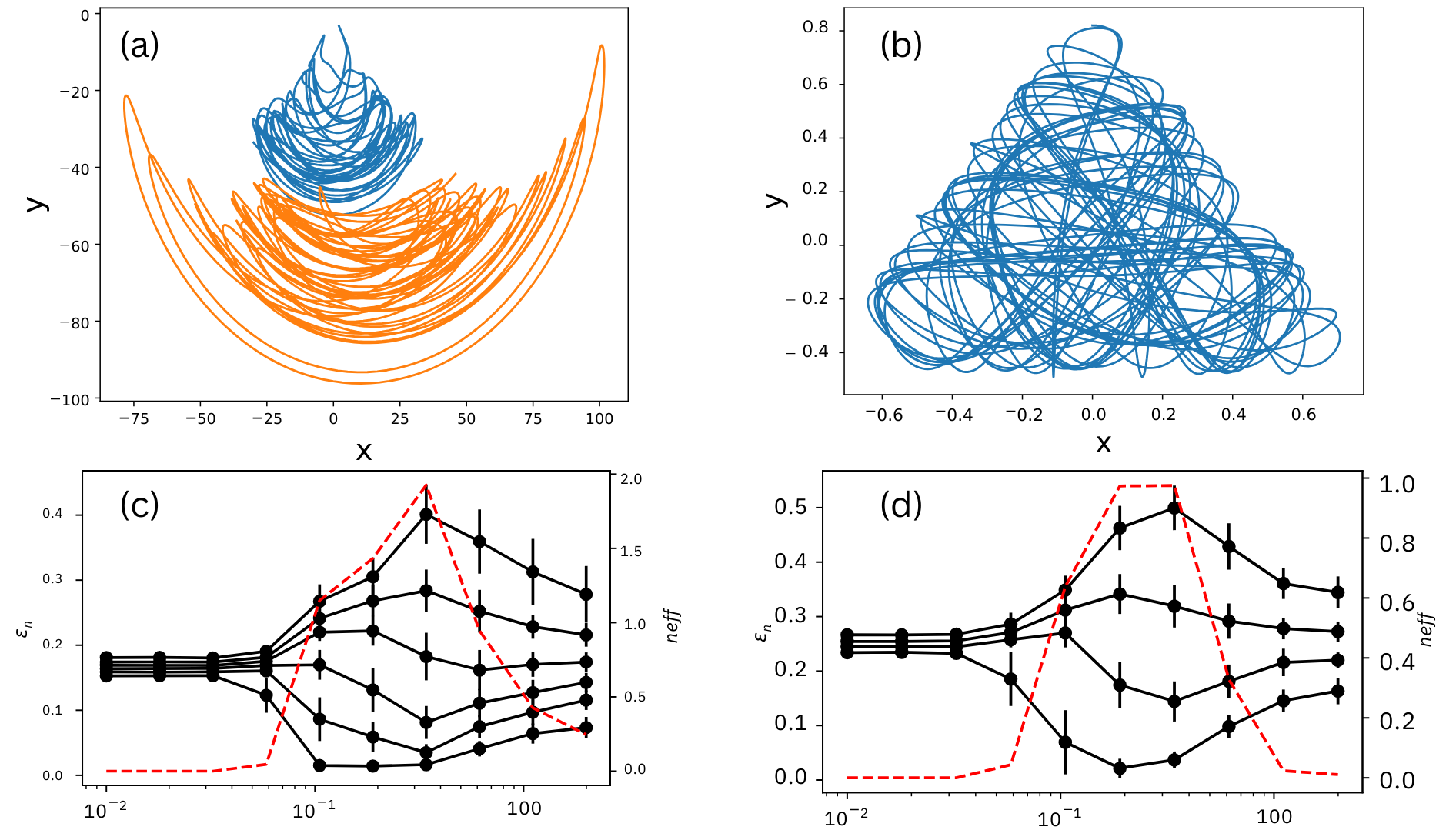}}
\caption{Trajectories of the test systems are shown in panels (a) and (b). Panel (a) depicts "Two bodies swinging on a rod," while panel (b) illustrates the "Henon-Heiles system." Panels (c) and (d) display the corresponding explained ratio diagrams, indicating the effective number of conserved quantities. The initial conditions and dimensionless parameters for system (a) are as follows: $\theta(0)=0.4$, $\phi(0)=0.41$, $\ell=0.05$, $R/L=0.01$, $m_1/m_2=1.1$ , with all time derivatives initially set to zero. The initial conditions of system (b) are: $x(0)=p_y(0)=0$, $y(0)=0.62$, $p_x(0)=0.1232$ } \label{fig5}
\end{figure}

Not only in complex systems covering all branches of physics \cite{c10,c12,c21,c1,c2}, but Even seemingly simple systems like the double pendulum, the Duffing oscillator, Chua's circuit, etc exhibit chaotic behavior under specific conditions \cite{c3,c4,c5,c6,c7,c8,c13,c14,c15}. Understanding and controlling chaotic dynamics, which are prevalent in everyday phenomena, hold practical significance. 

Traditionally, physicists have derived equations of motion, conservation laws, and symmetry conditions using a model-driven approach. They then analyze whether the system exhibits chaotic behavior and explore the conditions under which this occurs. This analysis often involves investigating Lyapunov exponents, which are key parameters for assessing chaos and sensitivity to initial conditions. Various computational methods exist for computing Lyapunov exponents, as well as for studying bifurcation diagrams and phase transitions that indicate when a system becomes chaotic. However, these methods typically require significant computational resources.
In contrast, this paper takes a different approach by analyzing chaotic systems using data-driven methods. Instead of relying on known equations of motion, we utilize phase space trajectory data as input for our analysis. This data-driven approach offers a novel perspective on understanding chaotic behavior and may provide insights that traditional model-driven approaches overlook.

Recent advancements in manifold learning, such as those seen in \cite{NEB}, have opened up new possibilities for analyzing phase space trajectories. One noteworthy development stemming from these advances is the AI Poincare algorithm \cite{AIP}. This algorithm is designed to identify and quantify conserved quantities within a system. In this manuscript, we intend to employ the AI Poincare algorithm to address three significant issues in chaos theory.

Firstly, we will utilize the algorithm to compute the Maximal Lyapunov exponent (MLE) more efficiently. Unlike traditional analysis methods that require numerous solutions within a very small initial conditions range \cite{c18,c19,c20}, the Monte-Carlo sampling \cite{NEB} enables us to achieve this with just two closely spaced initial conditions. This eliminates the need for extensive averaging, streamlining the computation process. 

Secondly, we will investigate phase transitions from regular to chaotic dynamics. This exploration will help us delineate the boundary between these states in dynamic systems, providing insights into the mechanisms underlying such transitions.
Lastly, we aim to address the measurement problem in chaos. This challenge involves reconciling theoretical predictions with experimental observations, which can be particularly difficult due to the complex relationship between motion and initial conditions or parameters in chaotic systems. Our approach involves identifying "integrable" and non-chaotic regions within overall chaotic dynamics, facilitating a more comprehensive understanding of their behavior that might make theoretical predictions possible on chaotic dynamics within those "integrable" regions in real-life phenomenons.

Through the application of the AI Poincare algorithm, our goal is to provide fresh insights and potential solutions to these enduring challenges in chaos theory. By doing so, we hope to contribute to a deeper understanding of complex dynamic systems and their behaviors.

The initial section offers a succinct introduction to the AI Poincare algorithm. Following this, the subsequent section delves into a novel approach for calculating the Maximal Lyapunov exponent, accompanied by an in-depth discussion highlighting its methodological advantages. Moving forward, the paper transitions to an examination of phase transitions from regular to chaotic dynamics. Finally, the last section focuses on identifying and analyzing "integrable" regions within chaotic dynamics. Notably, all assessments are conducted across three distinct dynamical systems: the double pendulum, two bodies swinging on a rod, and the Henon-Heiles system \cite{me,henon}.

\section{Overview of Monte-Carlo sampling and AI Poincare algorithms performance on test systems}

The AI Poincare algorithm is a machine learning tool that analyzes data from a system to determine how many conserved quantities the system possesses. It doesn't require prior knowledge of the system's equations. If the algorithm discovers that the number of conserved quantities is fewer than the system's degrees of freedom, it indicates that the system's behavior is chaotic. In simpler terms, AI Poincare helps us identify chaos in a system by analyzing phase space data.
AI Poincare consists of three steps, the first step is prewhitening which means standardization of the data and possible reduction of linear conserved quantities using the Principal Component Analysis (PCA) \cite{pca}. In effect, we are discarding linear invariants in the dataset so that the subsequent steps can focus on learning non-linear conserved quantities. The next step, referred to as "Monte-Carlo sampling," serves as the training phase. Following data scaling and application of the PCA , the algorithm proceeds by selecting a set of points on the manifold. It then introduces random perturbations to each point, guided by a specific length scale. Subsequently, the algorithm endeavors to revert each perturbed point to its original position. This iterative process aids in refining the algorithm's understanding of the underlying structure of the data manifold.
Once the Neural Network has completed its training, the algorithm becomes adept at accurately returning the perturbed points to their original positions on the manifold. It utilizes the RMS deviation as the loss function $Loss=something$ and employs the "Adam optimizer" \cite{adam} for dynamically optimizing the learning rate parameter. The neural network architecture includes two neural layers with 256 neurons each, the batch size is 1024 and initial learning rate is 0.01. Following the training phase, the algorithm perturbs several points on the manifold 2000 times, using training length scales, and subsequently restores them to their initial positions. As a result, the pulled-back points reside within the vicinity of the original chosen point, indicating that the neighbourhood effectively represents a tangent space of that point on the manifold.

Once local tangent spaces on the manifold are obtained, we can further process them by applying PCA to reduce their dimensionality. This reduction is based on a specific criterion: if the fraction of eigenvalues associated with corresponding eigenvectors of the space is significantly smaller than $\epsilon=0.01$, then the dimension of that local space is reduced. This reduction reflects the presence of conserved quantities, as conserved quantities tend to reduce the dimensionality of the manifold \cite{lan,lansx}. In simpler terms, this process of Monte Carlo sampling is equivalent to discovering the action-angle coordinates, where certain coordinates effectively represent the conserved quantities themselves.
The primary outcome of the analysis is encapsulated in the explained ratio diagram. This diagram illustrates the relationship between the dimensionality of the system and the number of conserved quantities. However, it's important to note that this diagram is effective only within a specific range of learning noise length scales, typically around $L=0.1$. Within this range, the diagram provides valuable insights into the dimensionality and the presence of conserved quantities within the system. 

We have applied these algorithms to two test systems "two bodies swinging on a rod" and "Henon-Heiles" systems. As you can see on Fig.(1) "two bodies swinging on a rod" system consists of a rope hanging on a rod with masses attached to its ends, the dynamics that the oscillation posses is an example of transient chaos and the lifetime of the oscillations sensitively depend on parameters and initial conditions. The "Henon-Heiles system" is a well known example of a non-linear chaotic dynamics and it is a model of non-linear motion of a star around a galactic center with the motion restricted to a plane. the dimensionless equations of motion for the "two bodies swinging on a rod" are the following \citep{me}

\begin{eqnarray}
(l\dot{\theta}^2-\ddot{l}+r\ddot{\theta})+\cos{\theta}=\frac{m_1}{m_2}\left[(\ell\dot{\phi}^2-\ddot{\ell}+r\ddot{\phi})+\cos{\phi}\right] \nonumber \\
\sin{\phi}=\dot{\phi}^2r-(\ddot{\phi}\ell+2\dot{\phi}\dot{\ell}), \quad
 \sin{\theta}=\dot{\theta}^2r-(\ddot{\theta}l+2\dot{\theta}\dot{\ell}) ~~~ \label{a11}
\end{eqnarray}

together with the inelasticity condition of the string
 
\begin{equation}
\label{a5}
 l_2=L-R(\pi-\phi-\theta)-l_1. 
 \end{equation}

and for the "Henon-Heiles system" the equations of motion describing the dynamics are following \citep{henon}

\begin{equation}
\begin{cases}
 \dot{x} = p_x \\
 \dot{p_x} = -x - 2xy \\
 \dot{y} = p_y \\
 \dot{p_y} = -y - (x^2 - y^2)
\end{cases}
\end{equation}

As you can see on Fig. (2) the panels (a) and (b) represent the trajectories of our test systems and below them, panels (c) and (d) correspond to their explained ratio diagrams, each black line in the plot shows the explained ratio of one phase-space dimension as a function of the random walk length scale during the training process. A small value of explained ratio can be interpreted as a detection of a conservation law. When the neural network is trained using very small walk length scales, it tends to learn the location of the specific points in the data set, but it is unable to learn about global shape of manifold and the estimated dimensionality is simply $1/N$, where $N$ is the number of phase-space dimensions. At length scales larger than the typical separation between points in the dataset, the training process fails to teach the neural network pull back the perturbed points correctly, and hence the inferred dimensionality is undefined. However, at intermediate walk length scales, the neural network learns the global shape of the manifold. For our test systems, since the "Two bodies swinging on a rod" is 6 dimensional in phase space and "Henon-Heiles" is 4 dimensonal in phase space, as you can see, the number of conserved quantities is not enough in either cases for the systems to be integrable, therefore the dynamics are chaotic.

\section{Calculation of the largest Lyapunov's exponent by Monte-Carlo sampling}

The theoretical basis of the "chaos parameter" lies in observing the dynamics within an $\epsilon$ radius ball of initial conditions in phase space. By monitoring the evolution of each principal axis of this ball over time, we can quantify the rate of change as the Lyapunov spectrum. However, direct observation of the $\epsilon$ radius ball entails significant computational demands. Consequently, in practical applications, determining whether a system is chaotic often hinges on evaluating the maximal Lyapunov exponent. This exponent signifies the logarithmic rate of difference between two radius vectors, where each radius vector denotes the distance of a phase space point from the origin. It's important to note that this evaluation typically involves multiple iterations for pairs of closely spaced initial conditions, followed by averaging the results.

\begin{equation}
\label{a12}
\lambda=\lim_{t\to t_0}~\lim_{\delta\phi_0\to 0} \frac{1}{t}\ln\frac{\delta\phi_t}{\delta\phi_0}
\end{equation}

In the context of developing a new data-driven method for analyzing chaotic systems, a slight shift in narrative is necessary. Traditional methods often rely on equations to solve for the behavior of systems with extremely close initial conditions or necessitate conducting two experiments with almost identical initial conditions. both scenarios work for the following method.
The proposed method operates as follows: Initially, two sets of phase space data, denoted as data1 and data2, are acquired for closely proximate initial conditions of the system. These datasets are structured as $mxn$ matrices, where 'm' signifies the number of experimental measurements, and 'n' represents the number of independent variables, such as position and momentum. Subsequently, the method involves computing the difference between data1 and data2, denoted as $data=data2-data1$. Finally, the obtained data is trained utilizing the previously outlined Monte Carlo sampling technique.
Following the training phase, the algorithm gains the ability to construct tangent spaces around the acquired data points. Subsequently, the PCA is applied to these tangent spaces, enabling the determination of eigenvalues for each eigenvector. Notably, the eigenvalues are directly linked to the variations along the corresponding principal axes. Importantly, the new dataset derived from the subtraction of two matrices, representing the phase space coordinates of the dynamics, plays a pivotal role in this process. The eigenvalues essentially quantify the average divergence between two solutions of differential equations, and since PCA is an orthogonal transformation the whole divergence can be calculated by Pythagorean theorem and therefore calculation of MLE is possible using the following equation 

\begin{equation}
    \lambda_{max}=\frac{1}{t}\ln\left(\frac{\sqrt{\Lambda_1+\Lambda_2+...+\Lambda_2n}}{\sqrt{\Lambda_01+\Lambda_02+...+\Lambda_02n}})\right. \label{ftle}
\end{equation}

where each $\Lambda_i$ corresponds to divergence between solutions on each i-th principal axis and $\Lambda_0i$ corresponds to divergence between solutions initially.

In Fig. (3), the variation of $\lambda$ over time is depicted, illustrating its progression through a plateau, this graph is constructed by the following steps: first the simulation period time should be divided into several n part, then for each calculation of MLE at each time period, the algorithm has to be trained on the data under $t+\delta t$ simulation time. Ultimately, it can be inferred that the Lyapunov exponent for the "Two bodies swinging on a rod" system locally attains a value of $\lambda=0.013$, while for the "Henon-Heiles" system, it amounts to $\lambda=0.083$. Notably, these values closely align with traditionally obtained results, affirming the efficacy of the proposed method. It's worth noting that traditionally, this result is obtained only after an averaging procedure that means calculation of Lyapunov's exponents for many different pairs of initial conditions on the same neighbourhood of point on the phase space and then averaging the results, which is sometimes tricky or even impossible especially for the systems that are characterized by transient chaos, because there might not be enough amount of pair initial conditions in the same neighbourhood of points that lasts enough amount of simulation time for data analysis due to the feature of transient chaos that systems dynamics might just end for various reasons, the clear example is "two bodies swinging on a rod". However this method only requires two solutions and it can be seen from the inset plot of the Fig.(3) that relative std(standard deviation) is close to zero, the relative std has been estimated by calculating the MLE using many different pairs of initial conditions within the same neighbourhood points and as you can see they do not vary for different pairs, meaning that there is no need for averaging procedure. The reason behind it lies in the mechanism of how "Monte-carlo sampling" works, it makes local tangent spaces by pulled back points, but these pulled back points corresponds to the solutions of other initial conditions, so some kind of averaging procedure is already encoded in the working mechanism of the algorithm.

\begin{figure}
\resizebox{\hsize}{!}{\includegraphics[width=10cm]{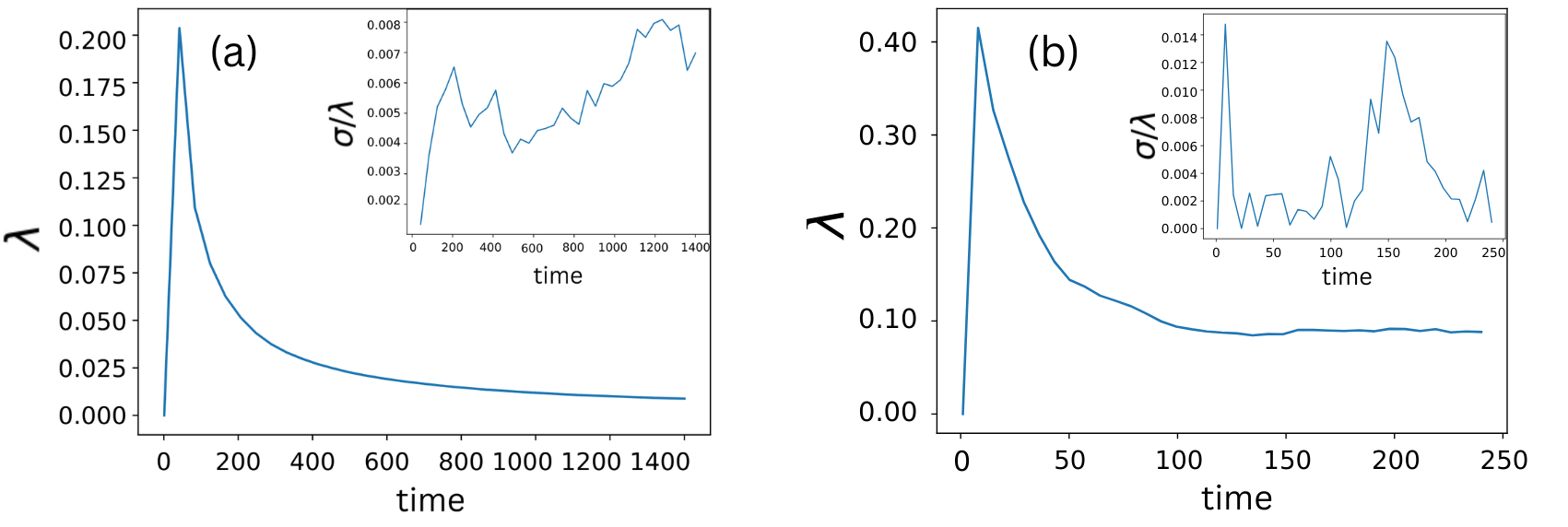}}
\caption{Lyapunov exponent results are shown. Panel (a) depicts "Two bodies swinging on a rod," while panel (b) illustrates the "Henon-Heiles system." The main plot describes the dependence of the Lyapunov exponent on time, and the inset plots show the relative standard deviations.} \label{fig5}
\end{figure}

\section{Phase transition}

Another crucial aspect of investigating chaotic dynamics involves determining the transition conditions for which parameters and initial conditions lead to chaotic behavior. Traditionally, this is accomplished through the use of bifurcation diagrams. However, in certain cases, such as the "two bodies swinging on a rod" system, constructing bifurcation diagrams presents challenges. For instance, in this system, swinging objects could potentially fall off the rod, abruptly terminating the dynamics. Moreover, the construction of bifurcation diagrams often requires extensive computational time for simulations, rendering it impractical in some scenarios.

\begin{figure}
\resizebox{\hsize}{!}{\includegraphics[width=10cm]{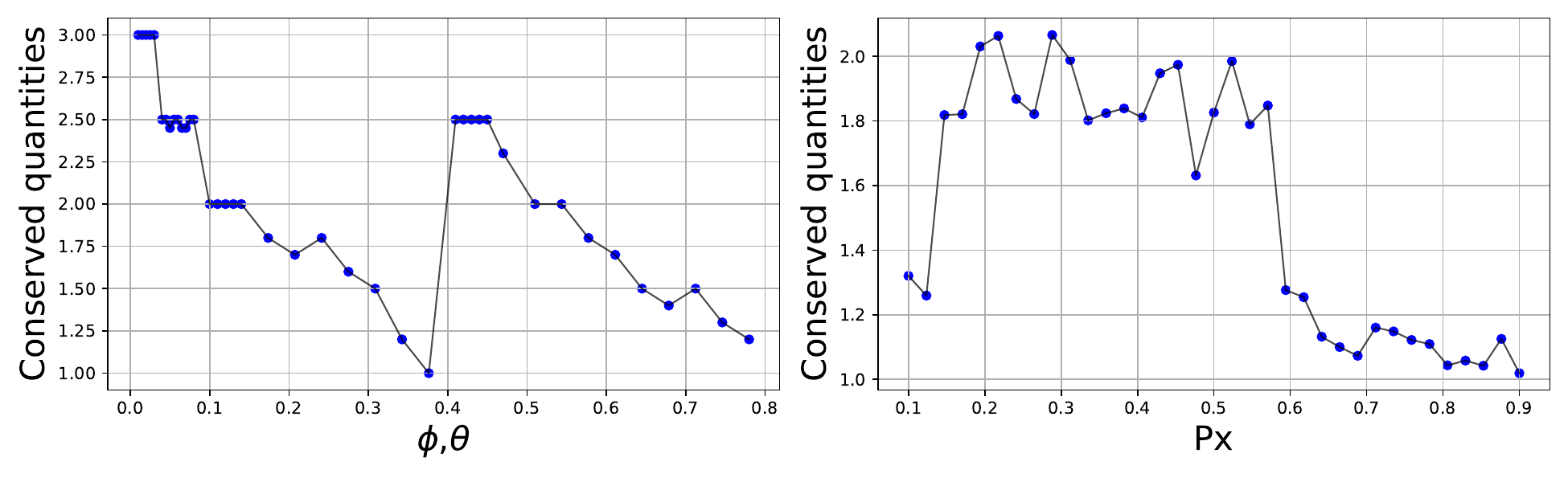}}
\caption{The graph shows how the effective number of conserved quantities depends on the initial conditions, indicating whether the systems are chaotic. Panel (a) corresponds to "Two bodies swinging on a rod," while panel (b) corresponds to the "Henon-Heiles system." } \label{fig5}
\end{figure}

However, the AI Poincare algorithm can be useful in this situation as well.
As we know, if the number of conserved quantities matches the number of degrees of freedom, the system is integrable. Therefore, we can create a graph illustrating the effective dimensionality of the manifold as a function of the parameter of interest. For example, if we seek to understand at which inclination angle the system becomes chaotic, we can systematically vary the angle and observe whether the system exhibits chaotic behavior. This observation involves tracking the effective dimensionality of the manifold, thereby allowing us to construct the graph illustrating the transition to chaos. In this article, we apply this approach to two previously mentioned systems, namely the "two bodies swinging on a rod" and "Henon-Heiles" systems. As depicted in Fig. (4), the system exhibits the integrable behavior, signified by a manifold effective dimensionality of 3, only for very small inclination angles. For the Henon-Heiles system, the degrees of freedom is 2 and for some initial conditions the system is integrable. Both of these findings align with existing knowledge, confirming the accuracy of our analysis as these systems have been extensively studied.

On both of the graphs, we observe the distinction between chaotic and regular periodic or quasi-periodic motions. However, it is noteworthy that the dimensionality of the manifold on these graphs is often represented by fractional numbers. While this may seem counterintuitive, it can be explained by the Monte Carlo sampling method employed. This method calculates the local dimensionality of the manifold at different points and then averages them. Consequently, the fractional effective dimensionality of the manifold arises from the fact that different parts of the manifold locally exhibit varying dimensions, which are subsequently averaged to produce the final result.

\section{Identification of "Integrable" regions in a chaotic trajectory}

As previously discussed in the preceding section regarding the fractional effective average dimensionality of the manifold, in this section, we introduce a novel approach to analyzing dynamics. Instead of solely focusing on calculating the average dimensionality of the manifold, we delve into observing the local dimensionality dynamics in each fractional average dimension scenario. This approach enables us to explore the intriguing aspects of the manifold that exhibit "locally integrable" characteristics. By examining these localized dynamics, we gain valuable insights into the behavior of the system on a finer scale, beyond the scope of traditional analyses.

Specifically, our approach involves observing a 2D section of the phase space. As an additional dimension on the graph, we represent the local dimensionality using different colours. This visualization technique enables us to elucidate the intricate structure of the manifold and discern regions with distinct local dimensionality behaviors, we also have made multiple 2D sections of phase space for better visualisation.

\begin{figure}
{\includegraphics[width=13cm]{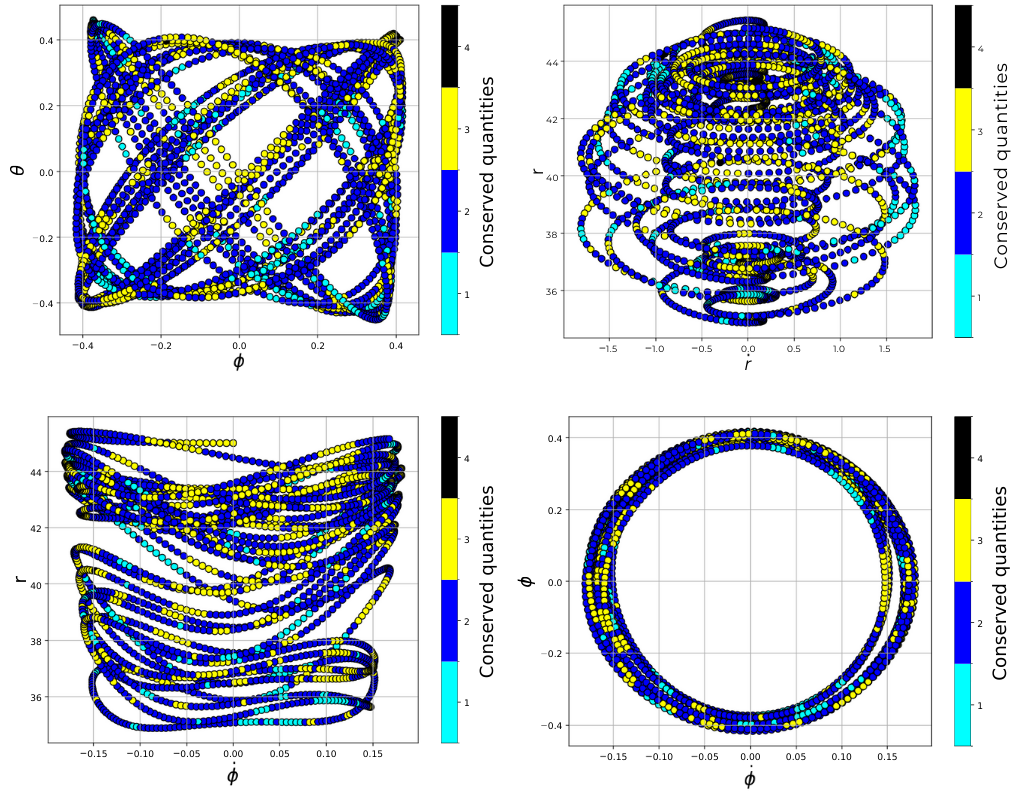}}
\caption{The graph shows the 2D phase-space planes of the system "Two bodies swinging on a rod," where each colour corresponds to local and different number of conserved quantities. The initial conditions and dimensionless parameters are: $\theta(0)=0.4$, $\phi(0)=0.41$, $\ell=0.05$, $R/L=0.01$, $m_1/m_2=1.1$ , with all time derivatives initially set to zero.} \label{fig1}
\end{figure}

\begin{figure}
{\includegraphics[width=13cm]{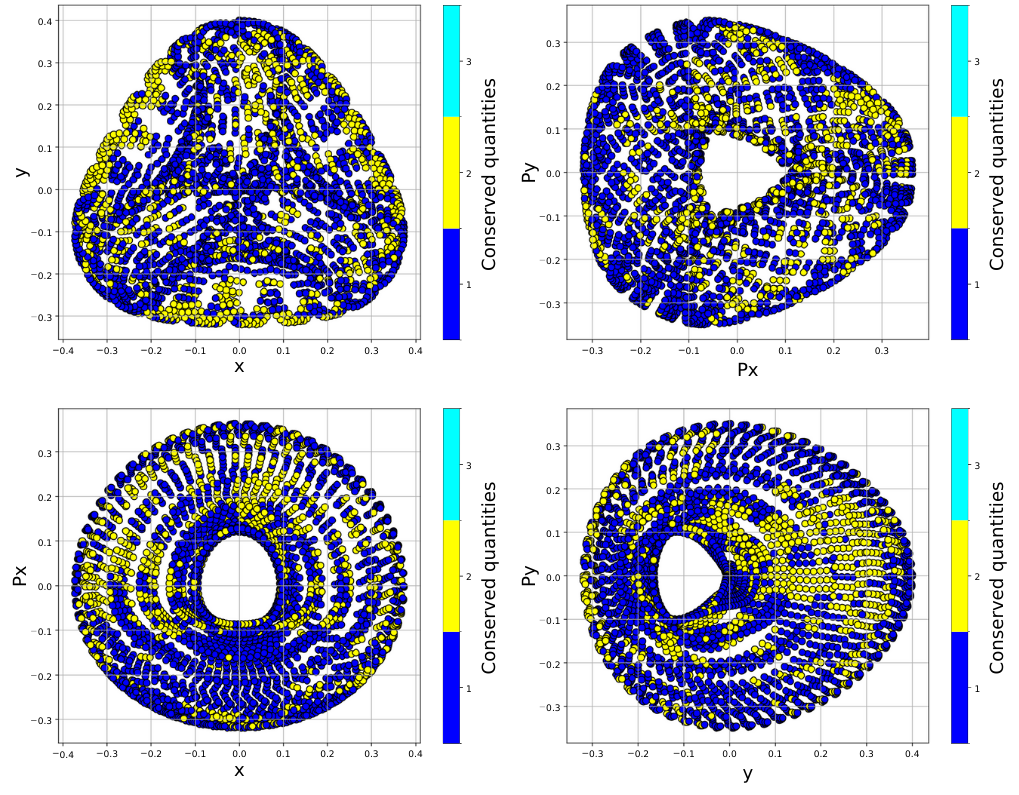}}
\caption{The graph shows the 2D phase-space planes of the system "Two bodies swinging on a rod," where each colour corresponds to local and different number of conserved quantities. The initial conditions and dimensionless parameters are:  $x(0)=p_y(0)=0$, $y(0)=0.62$, $p_x(0)=0.1232$.} \label{fig1}
\end{figure}

As depicted in Fig.(5) and Fig.(6) , which represents multiple 2D sections of the phase space, for the "two bodies swinging on a rod" and "Henon-Heiles system"

It is evident from these figures that multiple continuous regions within the overall chaotic dynamics exhibit integrable behavior in both of the test systems. In these regions, the effective number of conserved quantities equals the number of degrees of freedom, indicating integrability. the stability of the solutions in these regions have been tested and proved by numerical simulations. This observation provides valuable insight into the system's behavior, highlighting the presence of distinct regions where the dynamics are more predictable and less chaotic.

Indeed, the identification of such integrable regions within chaotic dynamics holds significant practical implications, particularly in real-life experiments and in controlling chaos. These regions offer potential opportunities for predicting system dynamics, even amidst chaotic behavior. By leveraging the existence of integrable regions, it becomes feasible to make accurate predictions about the system's behavior within these specific regions. This, in turn, opens avenues for controlling chaotic behaviors to some extent.

Practically, this means that in systems exhibiting chaotic dynamics, there may be certain regions where the behavior follows more predictable patterns akin to integrable systems. By identifying and understanding these regions, researchers can develop strategies to steer the system's behavior toward desired outcomes. This could have applications in various fields, including engineering, Astronomy, particle accelerator physics, and biology, where controlling chaotic phenomena can lead to advancements in technology, optimization of processes, and improved understanding of complex systems.

Indeed, while identifying integrable regions within chaotic dynamics holds promise for practical applications, it's crucial to acknowledge the significance of the size of these regions. If the regions are too small, their practical implications diminish. Therefore, it's imperative to ascertain how we can determine if a region is too small.

The answer lies in recognizing that even within chaotic dynamics, there exists a brief period of time during which theoretical predictions closely match experimental observations within an acceptable margin of error. However, this period is exceedingly short. It occurs because two solutions of a chaotic system require a small amount of time to diverge from each other by a distance greater than the margin of error. In other words, during this fleeting period, the discrepancy between theoretical predictions and experimental observations remains within an acceptable range.

We can estimate this small amount of time by considering the exponential divergence of trajectories in chaotic systems, often represented as $d_t=d_0e^{\lambda t}$, where $\lambda$ denotes the Lyapunov exponent. The Lyapunov exponent is inversely proportional to the characteristic time scale $T$ of divergence between trajectories, implying that $T \sim 1/\lambda$.

To define a "small amount of time" during which theoretical predictions align with experimental observations within an acceptable margin of error, we can set $T_s$ as an estimate. Typically, in chaotic dynamics, this alignment occurs within a margin of error range of $T_s=0.001/\lambda$ time units.

Thus, to ascertain the practical usefulness of integrable regions, we need to determine if these regions persist for durations significantly longer than $T_s$. If the duration of integrable regions exceeds $T_s$, they become practically useful for making accurate predictions and controlling chaotic behaviors. The continoues regions that last long enough to meet our "usefulness" criteria has been found in both of our test systems, it is indeed noteworthy to state that there are multiple integrable regions for each of our test systems.

\section{Conclusions}

In this article, we have introduced novel machine learning methods for analyzing chaotic dynamical systems, aiming to achieve several objectives. Our first goal was to introduce a new and relatively straightforward method for calculating the Lyapunov exponent, which only requires two closely related trajectory data points as input, unlike existing numerical methods. Subsequently, we investigated phase transition graphs and explored unique scenarios where conserved quantities were not whole numbers, labelling these trajectories as "almost integrable." Finally, we identified regions within overall chaotic trajectories that exhibited integrable behavior, which we termed "integrable regions." For both the calculation of the Lyapunov exponent and the identification of integrable regions, we also estimated the necessary duration for these regions to be practically useful.
These new methods were rigorously tested and applied to two dynamical systems: "Two objects moving on a rod" and the "Henon-Heiles" systems. We believe that the identification of integrable regions holds significant potential for controlling chaos across various fields, including engineering, technology, biology, astronomy, and accelerator physics. By leveraging these insights, researchers can develop strategies to predict and manage chaotic behaviors, leading to advancements in technology, optimization of processes, and deeper understanding of complex systems.

\bmhead{Acknowledgments}

The research was supported by the Knowledge Foundation of the Free University of Tbilisi. I would like to thank M.Dvalishvili for the great support, discussions, and memories.

\bmhead{Data Availability Statement}: The datasets generated during and/or analyzed during the current study are available from the corresponding author upon reasonable request.

\end{document}